# Photoemission and the Electronic Structure of PuCoGa$_5$

J.J. Joyce, J.M. Wills, T. Durakiewicz, M.T. Butterfield, E. Guziewicz, J.L. Sarrao, L.A. Morales and A.J. Arko

*Los Alamos National Laboratory, Los Alamos, NM 87545*

O. Eriksson

*Department of Physics, Uppsala University, Box 530, Sweden*

Abstract

The electronic structure of the first Pu-based superconductor PuCoGa$_5$ is explored using photoelectron spectroscopy and a novel theoretical scheme. Exceptional agreement between calculation and experiment defines a path forward for understanding electronic structure aspects of Pu-based materials. The photoemission results show two separate regions of 5f electron spectral intensity, one at the Fermi energy and another centered 1.2 eV below the Fermi level. The results for PuCoGa$_5$ clearly indicate 5f electron behavior on the threshold between localized and itinerant. Comparisons to delta phase Pu metal show a broader framework for understanding the fundamental electronic properties of the Pu 5f levels in general within two configurations, one localized and one itinerant.

PACS: 71.27.+a,71.28.+d,79.60.-i

The recent discovery of PuCoGa$_5$, the first Pu-based superconductor, with a T$_C$=18.5 K and unconventional superconductivity demonstrates the rich and complex nature of Pu-based materials[1]. The role of the 5f electrons in bonding and hybridization is intimately intertwined with the wide range of ground state properties found in actinide materials including enhanced mass, magnetism, superconductivity, as well as spin and charge density waves. Within the actinide series, Pu occupies the position between the clearly hybridized 5f states of uranium[2] and clearly localized 5f states of americium[3] which due to its J=0 ground state is a superconductor. Pu defines the localized/itinerant boundary for the 5f electrons in the actinides[4, 5]. There have been several recent papers reporting photoemission[6, 8, 7, 9] and electronic structure calculations[4, 5, 10, 11, 12, 13, 14] for the fcc ($\delta$) phase of Pu metal. A classic failure of density functional theory (DFT) within the local density approximation (LDA) or generalized gradient approximation (GGA) is observed in the case of $\delta$-Pu with the volume from LDA falling over 25% short (GGA only slightly better) of the experimental volume in the largest discrepancy to date in DFT for a crystal.

To accommodate the large volume of the delta phase of Pu metal, there have been magnetically ordered electronic structures proposed which, to some degree, effectively localize 5f electrons, reducing the contribution to the chemical bonding and increasing the volume[10], however there is no experimental evidence for magnetism in $\delta$-Pu[15]. Another approach used for $\delta$-Pu is the dynamical mean field theory (DMFT) which may offer promise with further development[11]. Currently the agreement between DMFT and experiment is weak with the PES shown in Ref.[11] being more of an idealized representation than actual data. The link between $\delta$-Pu metal and PuCoGa$_5$ is strong and follows the same path as the family of Ce-based heavy fermion-superconductors (CeMIn$_5$, M=Co, Rh, Ir)[16] having the cubic CeIn$_3$ as the root crystal structure. Here we have $\delta$-Pu as the parent cubic phase structure with Ga substituting on the face centers to form PuGa$_3$ and insertion of the CoGa$_2$ layer into the cubic structure to form PuCoGa$_5$ in the tetragonal phase. With the discovery of superconductivity in PuRhGa$_5$ [17] it does indeed seem reasonable that the PuMGa$_5$ family of compounds is following systematics much like the CeMIn$_5$ family of superconducting compounds.

Since the discovery of PuCoGa$_5$ there are currently three computational frameworks to describe the electronic structure of PuCoGa$_5$ nominally within the framework of DFT[18, 19, 20]. The theoretical efforts for PuCoGa$_5$ are following the path set forth in $\delta$-Pu, with conventional DFT pushing the 5f-electron intensity up against the Fermi level[18] or magnetic solutions[19] which serve to effectively localize some of the Pu 5f intensity below the Fermi level($E_F$). For PuCoGa$_5$, as in $\delta$-Pu, there is no experimental evidence for ordered magnetism. However, the magnetic solutions for both $\delta$-Pu and PuCoGa$_5$ provide valuable insight into the electronic structure. The agreement between photoelectron spectroscopy (PES) and calculation is improved in the magnetic calculations compared with conventional DFT calculations without magnetism. The magnetic state, which to some extent, effectively localizes f-electrons and removes them from the Fermi energy provides the needed spectral intensity away from the Fermi level to match the PES results[19, 10].

Understanding the nature of the 5f electrons is central to understanding the electronic properties of PuCoGa$_5$ as well as the broader Pu electronic structure problem. There is little photoemission data on metallic Pu systems beyond Pu itself with data for PuSe and PuSb appearing recently[21]. In this Letter we present experimental evidence for the Pu 5f electrons in two configurations, one well removed from $E_F$ and one directly at $E_F$ in a peak consistent with narrow band characteristics of other actinides[22]. Additionally, a mixed level model (MLM) calculation is presented showing remarkable agreement with the PES data where the Pu 5f electrons exist in two configurations, one localized and one itinerant[6, 13, 14]. The agreement between the model calculation and the PES data surpasses that found for the MLM and PES in δ-Pu[6, 14] which was already qualitatively the best agreement between calculation and PES (while preserving the known non-magnetic ground state).

Single crystals of multi-millimeter dimension PuCoGa$_5$ where grown by the flux growth method. The PES samples were from the same flux growth batch as samples described in the discovery paper which documents characterization[1]. The resistivity just above $T_C$ and the reproducibility of $T_C$ between single crystals[1] and polycrystals[17] indicate the high quality and robustness of the sample stoichiometry. Rectangular samples were selected for mounting and ordered along a low symmetry crystallographic direction. The analyzer acceptance angle was ±8° thus covering the reciprocal lattice at hv =40.8 eV and the data collection geometry was perpendicular to the sample surface. The sample surface was cleaned by laser ablation at a temperature (T) of 77 K in a vacuum of 6x10$^{-11}$ Torr. Laser ablation has been shown to be an excellent choice of surface preparation for actinides[7, 6]. There were no spectroscopic features identifiable as surface-related in the staged cleaning of the sample surface from native oxide through clean metallic surface. There was no sample surface contamination in the spectra with the 40.8 eV data having a sensitivity of about five percent of a monolayer for oxide-related degradation[8, 6]. The UV light source utilizes the He$_I$ (21.2 eV) He$_{IIα}$ (40.8 eV) and He $_{IIβ}$ (48.4 eV) lines. The energy resolution of the instrument (ΔE) was 60-120 meV with T=77 K and PuCoGa$_5$ in the normal state. With the three photon energies (hv ) one may obtain a qualitative assessment of the orbital cross-sections for the major electronic levels, the Pu 5f, Pu 6d and the Co 3d. The remaining levels (Ga 4s,4p, Co 4s, Pu 7s) show smaller cross sections over this energy range. Using the atomic cross-sections from Yeh & Lindau[23] as a qualitative guide one sees the Pu 5f and the Co 3d levels increasing from 21.2 eV through 48.4 eV while the Pu 6d level is decreasing. At hv =40.8 eV the cross section per electron for the Pu 5f, 6d and Co 3d levels are comparable and so this is a favorable energy to compare against calculation. The assumption of comparable cross-section at 40.8 eV has been explored for δ-Pu where the transition matrix elements were calculated at 40.8 eV and compared with the density of states showing only minor variations[14]. In addition to the matrix element calculations, the agreement between PES and MLM for δ-Pu in ref[14]. and here for PuCoGa$_5$, gives strong support for considering the PES spectra as representative of the electronic structure and not influenced in a substantive way by final state, satellite or screening effects not already accounted for in the model or processing of the data.

The basis of the MLM is a partitioning of the electron density into localized and delocalized parts, minimizing the total energy (including a correlation energy associated with localization) with respect to the partitioning[14]. The total energy of PuCoGa$_5$, and of δ-Pu becomes minimized for an atomic 5f$^4$ configuration with roughly one 5f-electron in a delocalized Bloch state. The four localized 5f electrons couple into a singlet, a many-body effect that is not accounted for within a single-particle picture. The hybridization between the conduction band states and the localized f-states (here a four electron multiplet state) is a key interaction needed to produce an accurate theoretical spectrum placing the MLM beyond the capacity of single-particle models such as LDA and GGA. The hybridization matrix element is calculated from the width of the 5f-resonance which is proportional to the 5f electron wavefunction at the muffin-tin sphere. Our value is consistent with the fact that spin- and orbital couplings within the 5f shell are important in this system. In addition to a broadening of the 5f-level, the hybridization matrix elements cause a reduction in the spectral weight. Hence the hybridization results in a broadening and a shift of the spectral weight. Figure 1

In Figure 1 we present the first photoemission data on $PuCoGa_5$. The spectrum was acquired at T=77 K, $\Delta E$=75 meV and hv =40.8 eV. The valence band spectrum in Fig. 1 contains three regions of interest; a sharp peak at $E_F$, a broad manifold centered 1.2 eV below $E_F$, and two small features at 5 and 7 eV below $E_F$, which most likely arise from the Ga and Co sp electrons. Along with the data we show lineshape analysis for the first 4 eV of the valence band. We use a Gaussian function for $\Delta E$, a Lorentzian for the photohole lifetime (natural linewidth), and a 77 K Fermi function. The fitting shows two distinct regions in the valence band, a narrow peak at $E_F$ with a width on the occupied side of $E_F$ of 100 meV and a broad manifold centered at -1.2 eV which is composed of two smaller peaks at -1 and -1.75 eV below $E_F$. The narrow peak near $E_F$ is directly cut by the Fermi function. The narrow peak at $E_F$ is very similar to such features found in several U compounds exhibiting narrow band behavior[22]. The fit to the data is good and the general description is consistent with that of δ-Pu[6, 8] showing two main regions of spectral intensity, both with substantial 5f contributions. In the case of $PuCoGa_5$ shown in Fig. 1 the main intensity around -1.2 eV is a superposition between Pu 5f and Co 3d states. Figure 2

The PES data and two theoretical calculations are shown in Figure 2. For cross-section reasons detailed above, the PES data was taken at hv =40.8 eV and compared to the calculations. The PES data are shown as diamonds, a GGA calculation as a dot-dash line and a calculation based on the mixed-level model shown as a dashed line. Like the lineshape analysis in Fig. 1, these calculations were Gaussian broadened for $\Delta E$, Lorentzian broadened for photohole lifetime (with $E^2$ dependence consistent with Fermi liquid theory) and the appropriate Fermi function for temperature. The above broadenings are not adjustable fitting parameters but rather fixed by known values of temperature, resolution and model lifetime. The dotted line is the MLM calculation without the broadenings. The bars are the nominal starting values for Pu 5f electron positions, the large bar for the localized manifold and the small bar the one 5f electron fully participating in the bonding. This delocalized 5f electron is part of the band structure calculation and fully hybridized with the conduction electrons with 5f character spread over an energy range of more than one eV. The PES data indicates that some significant fraction of this delocalized 5f weight resides with the $E_F$ peak. First, we note that the GGA calculation does not show particularly good agreement with the PES data, presenting the shortcomings for a GGA calculation with an open f-shell which concentrates the f-weight at $E_F$ (similar to the calculations of Ref.[18]). However, the mixed level model calculation shows remarkable agreement with the PES data. The calculation matches the data over the entire valence band region from the peak at the Fermi level through the main broad manifold at -1.2 eV to the two smaller features at 5 and 7 eV related to the sp electrons. Even the small shoulder in the PES data at -0.5 eV is reproduced in the calculation.

This level of agreement would be very good between PES data and a calculation for a simple metal or semiconductor; for a complex, ternary, Pu-based material it is exceptional. In particular, the peak at the Fermi level not only shows the correct energy and width, it also shows the correct ratio of spectral intensity to the main emission peak at -1.2 eV as well as the smaller features at 5 and 7 eV. The ratio of spectral intensities is important within the MLM framework in order to validate against experiment the fraction of localized vs. itinerant electrons in the calculation. In our calculations the total energy is minimized when four out of five 5f electrons are localized. Within the MLM computational framework, this represents a minimum in total energy compared with any other configuration (0, 1, 2, 3 or 5 Pu 5f electrons localized). This configuration also results in the correct volume for the unit cell, compared with the experimental lattice constant. This same solution was found for δ-Pu[13, 14] (see Fig. 2 Ref.[13]) and the plot of total energy vs. volume is very similar for δ-Pu and $PuCoGa_5$. The notion of the 5f electrons in Pu occurring in two different configurations for δ-Pu metal, one localized and one hybridized, was initially put forward independently on the experimental[7] and computational[13] fronts. With the results shown in Fig. 2 it would seem reasonable that this model for δ-Pu also works well for $PuCoGa_5$ and may have broad applications for Pu-based materials in general. Figure 3

In Fig. 3 we show the valence band PES data for $PuCoGa_5$ as a function of photon energy at 21.2, 40.8 and 48.4 eV. The main valence band feature (-1.2 eV) does not change width or shape substantially over this photon energy range. The peak at $E_F$ grows with respect to the main peak centered at -1.2 eV. The growth in intensity of the peak at $E_F$ combined with the narrowness of this spectral feature gives strong indication that this peak contains substantial 5f electron character. The itinerant nature of the peak cut by the Fermi energy is not in question but the extent of hybridization with conduction electrons is difficult to quantify. The existence of the peak at 21.2 eV, although small, probably indicates some non-f-electron character to the spectral feature. A narrow itinerant peak with substantial 5f character at the Fermi level is also consistent with the MLM calculation for $PuCoGa_5$.

The main peak in the PES spectrum at -1.2 eV is most likely a superposition of Pu 5f and Co 3d states. Both the Co 3d and Pu 5f cross-sections are increasing as a function of photon energy over the 21 to 41 eV interval, the Pu 5f cross-section increase by roughly a factor of 6 while the Co 3d cross-section doubles in value[23]. These differences account for the increase in the $E_F$ peak over the main peak since the $E_F$ peak contains primarily 5f character and the main peak a mix of 3d and 5f character. Also, the 5f character of the main peak would be necessitated by the relative intensities of the main peak and the Fermi level peak at 40.8 and 48.4 eV since a vast majority (over 90%) of the spectral weight is in the main peak it would be inconsistent to consider all of the 5f electrons in Pu to reside in the $E_F$ peak and the main peak to consist exclusively of Co 3d states. The composition of the main peak being a mixture of Co 3d and localized Pu 5f electrons is consistent with the MLM calculation for $PuCoGa_5$ which shows the strongest Co 3d intensity in the energy interval -0.75 to -2 eV.

In summary, the PES data for $PuCoGa_5$ gives solid indication of the Pu 5f electrons occurring in two configurations, one itinerant and being located at the Fermi level, the other localized and centered 1.2 eV below $E_F$. The MLM calculation shows this same Pu 5f arrangement, with one 5f electron being itinerant and the other four 5f electrons being localized. The agreement between PES and the MLM for $PuCoGa_5$ is impressive. This electronic structure arrangement with the 5f electrons in a dual mode is also found in δ-Pu metal[6, 14]. With reliable PES data that are in good agreement with theoretical calculations, which also give good volumes and the correct magnetic configuration, a path forward for understanding the complex electronic structure of Pu-based materials is at hand.

Acknowledgments Work performed under the auspices of The U.S. Department of Energy.

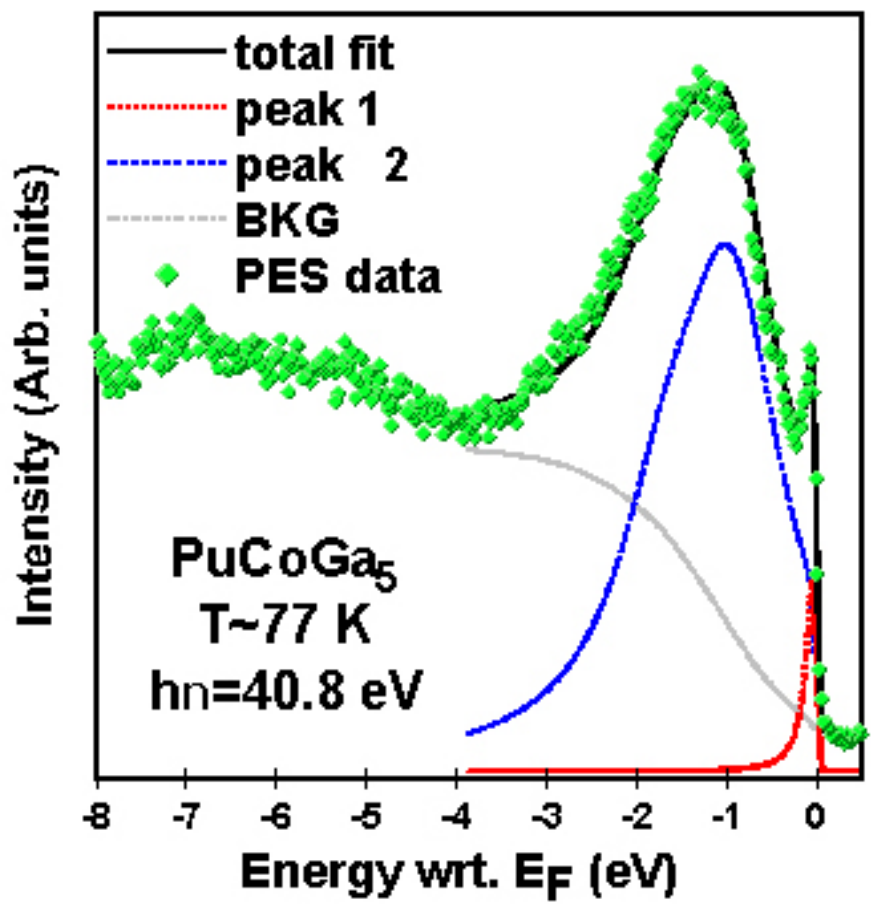

[Figure 1: The photoemission spectrum for PuCoGa$_5$ at hν =40.8 eV and T=77 K is shown with ΔE=75 meV. The data show a narrow feature at $E_F$, a broad manifold 1.2 eV lower, and two smaller features at 5 and 7 eV below $E_F$. Also shown is the lineshape analysis with a narrow peak cut by the Fermi function, and a broad feature centered on 1.2 eV made up of two components. (Data as diamonds, total fit solid line, Fermi level peak dotted line, main peak, dashed line and background dot-dash line.)]

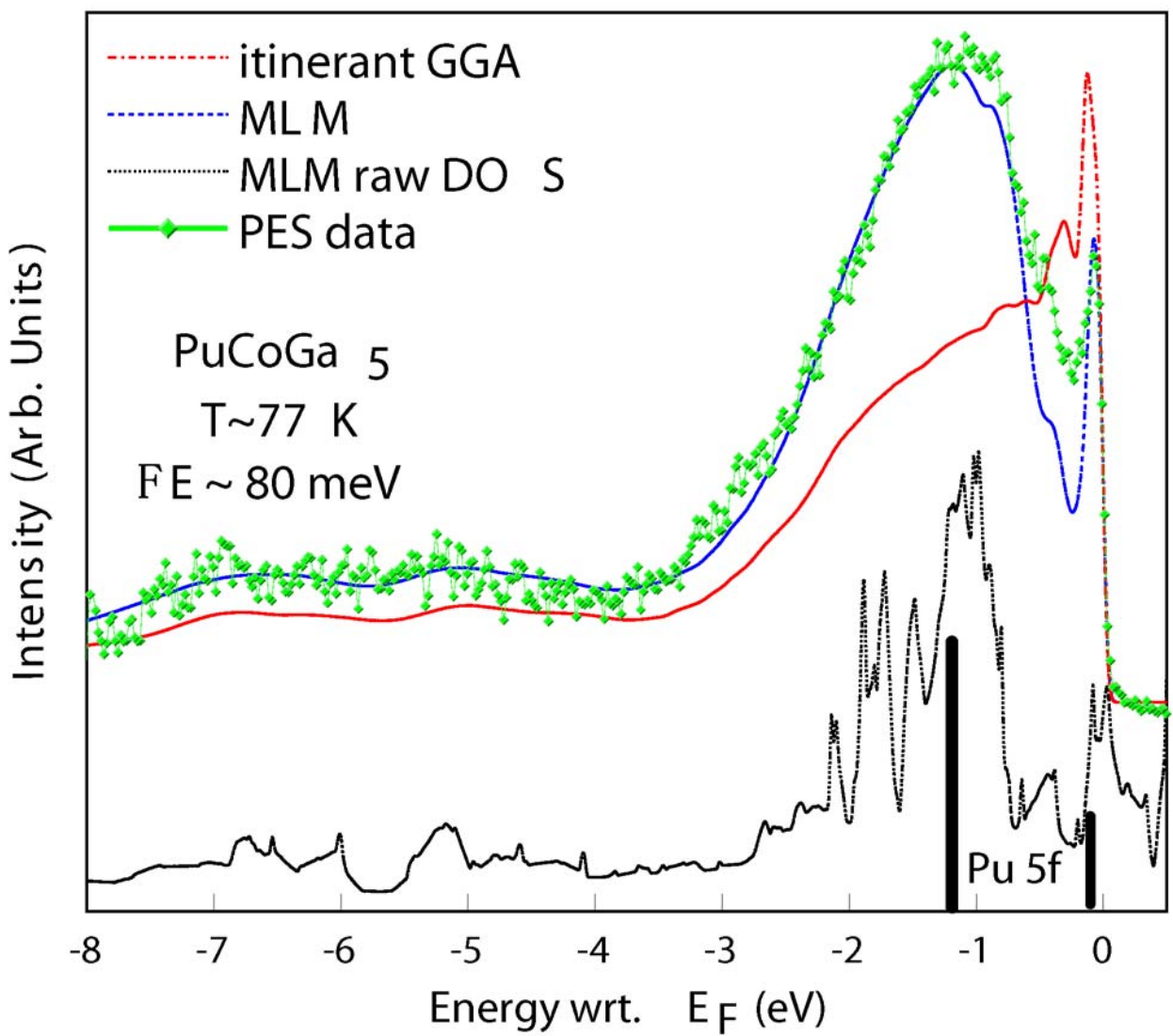

[Figure 2: The photoemission data for PuCoGa$_5$ is compared against model calculations. The experimental valence band PES data is shown as diamonds, a GGA calculation as a dot-dash line, a MLM calculation as a dashed line, and a raw MLM DOS as a dotted line (f-electron postion centers represented by solid bars). The calculations (less the raw DOS) have been processed for photohole lifetime, instrument and Fermi function broadening. The agreement between MLM and the PES is excellent.]

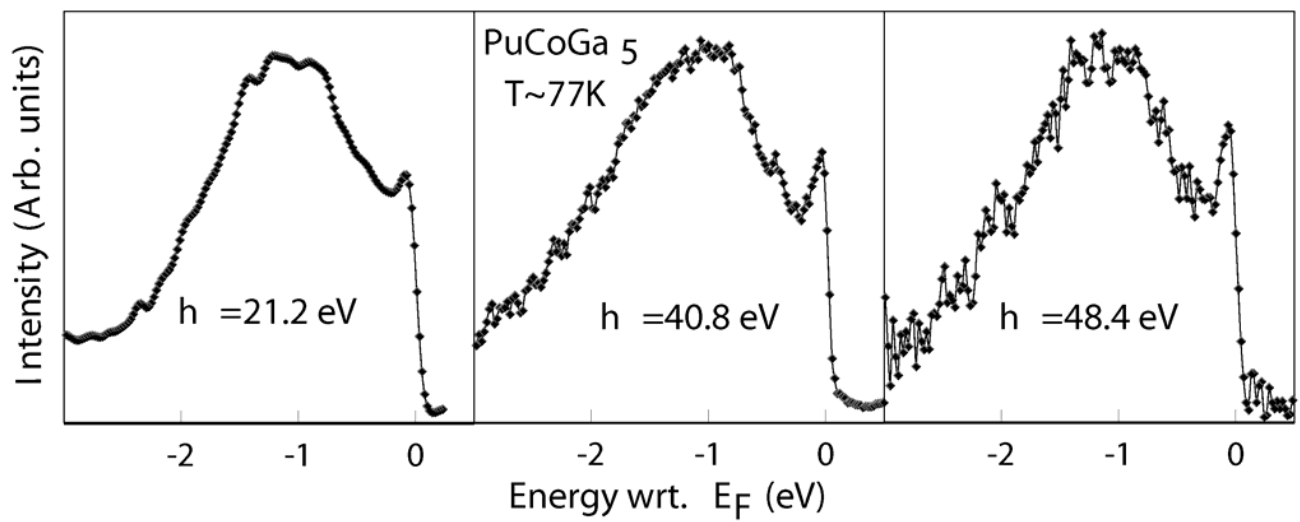

[Figure 3:Photoemission data taken with photon energies of 21.2, 40.8 and 48.4 eV. There is an increase in intensity of the peak at $E_F$ with respect to the main spectral intensity at 1.2 eV consistent with the peak at $E_F$ containing substantial but not exclusive Pu 5f character while the main peak centered at -1.2 eV is characterized by Pu 5f as well as Co 3d intensity.]